COSMIC COINCIDENCES: ANOTHER EXPLANATION.

by Helier Robinson.

There are a number of parameters in theoretical physics which are only known empirically: there is no theoretical basis for them having the values that they do. Among these are the rest-masses of the various wave-particles and the ratios of the strengths of the four fundamental forces. All of these parameters could have different values, according to theory, but if they did then life in our Universe would be impossible. The fact that they all do have values that allow life is called the problem of cosmic coincidences, also known as the fine-tuning problem. Each of these actual values is quite improbable, and the totality of them together vastly more so: Lee Smolin[1] has calculated that the probability of this is about one chance in $10^{229}$; and he lists four proposed explanations of this extraordinarily improbable fact.

The first of these is called the anthropic explanation. It supposes that in the one Universe there are a very large number of mutually isolated mini-universes, each taking on its own values of the parameters, at random. Given enough of these mini-universes, at least one will have the right values of its parameters so that stars, galaxies, and life are possible. Our own mini-universe has stars, galaxies, and life, because we exist, and so is one of these very improbable ones. The main objection to this is that it multiplies universes far beyond necessity, contrary to the principle of parsimony of hypothesis; also known as Occam's Razor, which says "Do not multiply entities beyond necessity." In contemporary parlance this means that one should not invent more theoretical entities than are needed to explain the empirical data.

A second explanation is that these cosmic coincidences are a "fine-tuning" of the Universe, and fine-tuning implies a Tuner, or God. The objection to this explanation is that, to paraphrase Laplace, physics has no need of that hypothesis. This is because it is, in principle, neither falsifiable nor verifiable; and because, as an explanation it merely pushes the existence problem back a stage, by raising the question as to why there should be a Tuner.

A third is the claim that in the ultimate analysis there is only one mathematical system that is consistent; since our Universe is mathematical and consistent, and contains life, it follows that this one consistent mathematical system must have these values of the parameters. Not only does this seem very implausible, but how could the uniqueness of this one consistent system be proved?

The fourth explanation is Smolin's own[2]. He suggests that when a black hole shrinks towards a singularity it becomes like the supposed singularity out of which the Big Bang grew, so that every time a black hole forms a new Big Bang universe is born, in a space-time distinct from our own. Black holes form with the collapse of large stars, so their formation requires universes with stars, hence galaxies and life, so the majority of new universes will come from black holes in universes like our own. Lee Smolin claims that this theory is in principle testable, unlike the other three: a most important point for a scientific hypothesis. However, it does have the same objection as the anthropic explanation: far too many mini-universes.

A fifth explanation offered here is due to that genius Gottfried Wilhelm Leibniz (1646-1716), who said that the actual world exists necessarily because it is the best of all possibles. He was widely misunderstood on this because everyone assumed that he was referring to the world

COSMIC COINCIDENCES: ANOTHER EXPLANATION.

of empirical phenomena that we all experience around us. But he was not: he was referring to the world of underlying causes of these empirical phenomena — the underlying causes that scientists try to describe, frequently successfully, with mathematical theories. I will refer to this as the *underlying world*, as opposed to the empirical world of everyday experience. Obviously the best of all possible worlds would be less than the best if it did not have stars, galaxies, and life. But it does seem at first that otherwise this explanation is little better that the Tuner one — except for one thing: there is an intriguing argument that the actual world, among all possible worlds, is the best because *the best exists necessarily*. What follows is an exposition of this argument. It is based on the concept of necessary existence, or actuality; if this concept is neither self-contradictory nor implies a contradiction then whatever has this necessary existence must exist in at least one among all possible worlds, making that one possible world actual.

We have to digress a little first. The approach used here arose from developing a philosophy of mathematics based on relations rather than sets, on the ground that in set theory the definition of relations involves a vicious circle: the definition of relations as subsets of Cartesian products presupposes the relation of set membership and the relations of logical argument forms. Since relations, arguably, are more important than sets in mathematics, and sets can be defined easily in terms of relations, this meant making relations fundamental in mathematics. This in turn led to distinguishing three kinds of defined sets, called intensional, extensional, and nominal sets; the intension of a set, if it has one, is all those properties that all and only its members possess; the extension of a set is the totality of its members; and a nominal set is a set that has a name or description but does not otherwise exist. From this there developed three kinds of mathematical meaning, called intensional, extensional and nominal meaning. It turns out that only nominal meaning can have, or lead to, contradictions and paradoxes, while only intensional meaning can have axiom generosity (that is, lots of theorems emerging out of the axioms), and extensional meaning has neither contradictions nor generosity. Intensional meaning is essentially relational: relations and their properties constitute intensional meanings. So if mathematics is to be rich and consistent it must be confined to intensional meaning. In particular, two concepts in common use can be shown to have no intensional meaning: *infinity*, and *chance*. (I must apologize for giving little detail about all this because of lack of space; anyone who is sufficiently interested in what follows can easily get more on the web[3].) In turn, this led to the question of mathematical physics, and whether the reality it describes — the underlying world — is relational. So we next consider the properties of relations in general.

1. The defining property of a relation is that it has terms, which are what it relates.
2. All and only relations have terms. The number of terms that a relation has is called its *adicity*. Adicity, in intensional mathematics, is the origin of number.
3. We can see relations but cannot say what they look like; we can touch them, but cannot say what they feel like; similarly with hearing them, tasting them and smelling them. For example, we can see that his hat is on his head, but what does *on* look like? And we can hear that this note is higher than that, but what does *higher than* sound like? This is because all these sensations are concrete, and

        relations have no concrete properties: they are abstract entities, but none the less usually real — real in the sense of existing independently of being perceived.

4. Relations cannot exist if their terms do not exist, and if their terms do exist the terms must be appropriately arranged for the relations to exist. When arrangements of terms change so as to have a relation come into existence, the relation is said to *emerge*, or to be *emergent*, out of those terms and that arrangement, and if further change of arrangement causes the relation to go out of existence, it is said to *submerge*. (Arrangements of terms consist of all the relations between these terms.)

5. Besides having terms, relations may be terms of other relations.

6. Relations have intrinsic properties, by which they can be distinguished. They also have two kinds of extrinsic properties: their terms, and the other relations of which they are terms; the latter are called their *upper extrinsic properties*. Their intrinsic properties determine the *kind* of relation they are, and their terms determine the *instance* of that kind.

7. Relations form what might be called *natural sets*: the sets of their properties, the sets of their terms, and the sets of their upper extrinsic properties.

8. Relations have a value, called their *hekergy*. If the number of arrangements of the terms of a relation with which the relation emerges is $e$ and the total possible number of arrangements of the terms is $t$ then the hekergy of that relation is $\ln(t/e)$. Hekergy is a generalization, to relations, of negative entropy[*].

9. A *structure* is an arrangement of relations relating relations which relate relations, to various levels. The lowest level of these relations is called the *prime level* of the structure. Relations at the prime level are called *separators*; they are characterized by being both terms and relations of these terms. If you think of a closed line of hyphens, then each hyphen is both a term of two hyphens, and relates two hyphens. That is, each hyphen separates two hyphens and has two hyphens as upper extrinsic properties. Separators might be spatial separators and temporal separators — Planck lengths and Planck times, perhaps — and causal separators. The prime level of a structure is known as level one in that structure. Relations emergent from separators are relations at level two; relations emergent from these are at level three, and so on. For example, a prime level might contain separators that are indivisible lengths and separators that are right-angle separators; four angle separators, each separating two lengths, and four lengths, each separating two right-angle separators, would form a square. This would be a level two structure having the emergent property of an area. And six such squares could form a level three structure: each square separating two pairs of squares through two right angles, so as to form a structure of squares called a cube, having the

---

[*]This is a misnomer: there is no such thing as literally negative entropy, since there is no negative temperature on the Kelvin scale; the term should be *negated entropy*.

        emergent property of a volume. In general (with two exceptions) each level consists of structures of lower level structures and various combinations of separators between them. The exceptions are the prime level and the top level (see 12).

10. A structure which has many levels of emergence is said to have *cascading emergence* out of its prime level. An example of cascading emergence is: wave-particles, atoms, molecules, cells, animals and plants, societies of these, ecosystems, and the biosphere. Another example is axiom generosity: the cascading series of definitions and theorems in mathematics, emergent out of an axiom set, such as Euclid's axioms, out of which emerge all of Euclidean geometry; or the concept of a right triangle, out of which emerges Pythagoras' theorem and all of trigonometry. Because a structure consists of two or more substructures, the number of structures in a level is smaller that the number of structures in the next level below. Thus there are a fewer number of atoms than there are numbers of protons or of neutrons or of electrons; a fewer number of molecules than the number of atoms, and so on. Conversely, as you go to higher levels the variety of structures increases: thus there is a greater variety of molecules than the variety of atoms.

11. At various levels of a structure a relation may emerge that has a property that does not emerge at any lower level. This property is novel at this level; it is called an *emergent property* and the lowest level at which it emerges is its *emergent level*. Examples of emergent relations having novel properties are the working order of a simple machine, emergent out of a correct arrangement of its parts (which are not themselves machines); the specific function of a particular kind of knot, emergent out of an arrangement of loops and threadings of a cord (a knitted sweater is a complex knot, having the emergent property of keeping you warm); a melody, emergent out of an arrangement of notes; life, emergent out of arrangements of molecules; and mind, emergent out of brain.

We now introduce two terms: *possible* and *actual*; the possible is defined in paragraph 16 below, and the actual in paragraph 21.

12. An intensional structure which is complete is called a *possible intensional world*, or *possible world* for short. Because a possible world cannot be infinite, it must have a finite prime level; and because the number of structures diminishes with height of level, it must have a highest level, called its *top level*. There is only one relation at the top level, since if there were more that one, these could be related by one relation, and this latter would be the top relation.

13. The variety of possible worlds is huge. It comes from the possible variety of prime levels, which comes from the possible variety, and numbers thereof, of prime relations, or separators. All these possible worlds are, of course, exclusively intensional: they consist only of relations and their properties.

14. The *value* of a possible world is the sum of the hekergies of all of the emergent relations in that complete structure, divided by the number of separators in its prime level. It is assumed that this value is proportional to the height of the top level of that world; if this should be incorrect then the value of a possible world is the height of its top level.
15. The best of all possible worlds is the one that has the greatest value; this world is called **G**. As we shall see below (28) it has a higher top level than any other possible world. Its top relation is called **T**.
16. There are two kinds of mathematical existence: possible existence and actual existence; this is because although all possible worlds have possible existence, only one can have actual existence, as we shall see. Possible existence is logical consistency: both intrinsic and extrinsic. Actual existence is explained in 21 below.
17. Necessary mathematical existence is existence which is necessitated. There are three kinds of extrinsic necessary existence and one kind of intrinsic necessary existence.
18. Of extrinsic necessary existence there are, first, causal and logical necessities, which are relations of singular possibility: the antecedent of such a causal relation is the cause of the consequent because if the antecedent exists then the existence of the consequent is the only possibility; and similarly for logically valid antecedent and consequent. These are called *causal necessary existence* and *logical necessary existence*.
19. Second, there is *top-down necessary existence*: because a relation cannot exist if its terms do not exist, the existence of a relation necessitates the existence of its terms.
20. Third, there is bottom-up necessary existence: if the terms of a relation exist and are arranged so that the relation emerges into existence, then that emergence is necessary.
21. Fourth, there is the possibility of a relation having *intrinsic necessary existence*, or I.N.E. for short. Such a relation would exist necessarily, by its own intrinsic property. This existence is actual existence, the source of all actuality. Note that because the concept of I.N.E. is not self-contradictory it must, like all intrinsically consistent concepts, exist in at least one of all possible worlds.
22. A possible world has *circular self-necessitation* in that the existence of its prime level bottom-up-necessitates the existence of all of its cascadingly emergent relations, including its top relation; and this top relation top-down-necessitates the existence of all lower level relations, including the prime level.
23. This circular self-necessitation is possible circular self-necessitation; every possible world possesses it. If a possible world has a relation that has I.N.E. then the circular self-necessitation is necessary, or actual, circular self-necessitation. Such a world is thereby an actual world.

From this point on the argument may be summarized as follows: (i) at most one possible world can be actual; (ii) at least one possible world must be actual; (iii) at most one possible world containing a relation possessing I.N.E. can exist; (iv) at least one relation possessing I.N.E. must exist; and (v) the only way in which all these four conditions can occur together is if this relation possessing I.N.E. emerges as the top relation of a possible world having a higher top level than any other possible world, which makes the relation possessing I.N.E. unique among all possible worlds; and the possible world possessing this unique relation must be the best of all possibles, **G**; and this unique relation must be **T**, possessing the emergent property of I.N.E. Thus:

24. If two possible worlds were actual then the two could be related into one, in which case neither would be complete (see 12), hence not a possible world. So at most one possible world can be actual.
25. Some relations are actual (because cascading generosity exists in mathematics and because of emergent relations in the underlying world) and so part of a possible world. So at least one possible world is actual.
26. If more than one relation possessing I.N.E. were actual, in different possible worlds then two or more possible worlds would be actual, which is impossible, by 24. So at most one possible world possessing a relation possessing I.N.E. can be actual. And if two relations possessing I.N.E. were actual in one possible world then one of them would have to have an emergent level lower than the top level of that world (see 28 below), in which case such a relation could emerge at that level in another possible world, and make that second possible world actual — which is impossible.
27. The one possible world that is actual has nothing actual outside of it, so its actuality cannot be externally caused; and it cannot be caused by chance because there is no intensional chance. So it must be internally caused — self-caused, by possessing a relation that has I.N.E.
28. The only way in which all four of these conditions can be met is that both (i) the actual possible world has a top relation having I.N.E., and (ii) the top level of this possible world is both higher than the top level of any other possible world and is the emergent level of I.N.E.
29. Since the number of emergent structures in a possible world decreases with level, it follows that the possible world that has the highest top level also has the largest number of separators in its prime level and so the largest number of structures in subsequent levels, and so is both actual and is the best of all possibles. So the actual world is **G** and its top relation, **T**, is the one and only relation possessing I.N.E. In other words, **G** is the best of all possible intensional, or relational, systems and is actual because it is the best.

(It might be argued that there is no difficulty in defining a prime level much greater than that of **G**, and that this would thereby have a higher top level. To show the fallacy of this, consider a parallel case: it was claimed above that there is no intensional meaning to infinity, from which it follows that there must be a greatest cardinal number, which we might call *g*. We can then define a number, *g+n*, where *n* is any number, and prove thereby that *g* is not the greatest cardinal. However, to define something does not mean that it actually exists, and if it does not exist then the definition has nominal meaning only — it has no intensional meaning. In other words, *g* is the greatest *intensional* number. Similarly, **G** is the intensional possible world with the highest top level; anything else that is claimed to be higher has only nominal meaning.)

We next consider the relationship between **G** and the underlying world, the world of underlying causes of empirical phenomena, the world that theoretical physics tries to discover. Let us name the underlying world **U**, and ask how it is related to **G**. We have already had an indication of this in 10, above: both **U** and **G** have cascading emergence. But we can do better than an indication.

We begin with the question: how can we explain the fact that mathematical theories predict empirical novelties, successfully and often? There is only one non-trivial answer[**]. Predictions of empirical novelty are logically necessitated by the mathematical theory, and the predictions come true because the novel phenomena are causally necessitated by the underlying world. If the predictions were not logically necessitated then they would be mere guesswork; and if the novel phenomena were not causally necessitated then they would happen only by chance. Guesswork and chance could conceivably combine successfully on rare occasions, but the sheer quantity of successful predictions of empirical novelty in the history of mathematical physics rules this out. So we can say with assurance that mathematical physics contains relations of logical necessity (that is, mathematical necessity), and the underlying world contains causal relations, some of which latter cause empirical phenomena. Not only that, but the antecedents of the logical necessities must describe essential features of the causes, and the consequents of the logical necessities must describe essential features of the effects — that is, of the empirical novelties. So in short, mathematical theories predict empirical novelties, successfully and often, because they are true. They are true in the sense of correctly mapping, mathematically, essential features of the underlying world — true at least in so far as they predict successfully.

We next note that a logical necessity is a relation between antecedent and consequent, and a causal relation is a relation between cause and effect. So mathematical physics, in so far as it is true, is a part of **G** — since **G** is the only possible actual intensional world — and at least some of **U** is correctly described by **G**. So some of **U** must be a structure of relations, and that structure must be a part of **G** — since **G** is the only possible actual intensional world. So both true mathematical physics and some of **U** are parts of the one actual structure **G**. Consequently **G** and **U** must be identical, one and the same: because if they were similar, but two, there would be two possible worlds, which is impossible. Thus when we say, in ordinary language, that the world we live in is the actual world (as opposed to any other possible world), this word actual has the same meaning as the actual that is derivative from I.N.E. Hence **U** exists necessarily because it is the best of all possible worlds — as Leibniz claimed.

---

[**]A trivial answer is: the predictions come true because Satan is trying to deceive us.

So, is there any other physical evidence, besides the cosmic coincidences, that the underlying world of theoretical physics is the best of all possibles? Successful prediction of empirical novelty is partial evidence. And there are at least three other items. One is the fact that spiral galaxies are in a state far removed from equilibrium[4], and remain so for billions of years, which enables them to make new stars and planets, and hence life. A state far removed from equilibrium is one of very low entropy, hence high hekergy, and is to be expected in the best of all possible worlds. The second item is the importance of symmetries in theoretical physics; a symmetry is a relation that has hekergy, a hekergy that its absence does not have, and so is to be expected in a world of high hekergy. The third item is the effectiveness of stationary principles, such as paths of least action and shortest time; such a path has no equivalent paths, so the number of possible arrangements of terms, $e$, is one and the hekergy of that path is maximal. Indeed, the actual world being the best of all possibles is the ultimate stationary principle.

On the other hand, this best of all possible worlds is deterministic; how does this mesh with the indeterminism of quantum mechanics? I suggest that this indeterminism is due to quantum mechanics being incomplete, and that completion requires top-down necessity. So how the wave function collapses is top down necessitated.

Thus Leibniz's bold statement that the actual world is the best of all possible worlds can be proven to be true, via a melding of mathematics, physics, and unfashionable philosophy. And the seemingly improbable factors that come together to allow life to exist in the universe are in fact dictated by the truth that the best possible Universe exists necessarily.